\documentclass{article}

\usepackage{amsfonts, amssymb, amsthm}

\usepackage{epsf}  
\usepackage{graphicx}
\usepackage{subfigure}
\usepackage{pstricks}
\usepackage{pst-node}

\begin{document}

\def\H{\mathcal{H}}
\def\p{\partial}

\def\N{\mathbb N}
\def\C{\mathbb C}
\def\R{\mathbb R}
\def\Z{\mathbb Z}

\def\be{\begin{equation}}
\def\ee{\end{equation}}
\def\bea{\begin{eqnarray}}
\def\eea{\end{eqnarray}}

\def\w{\widetilde}

\def\a{\alpha}
\def\b{\beta}
\def\d{\delta}
\def\e{\epsilon}
\def\g{\gamma}
\def\k{\kappa}
\def\l{\lambda}
\def\la{\Lambda}
\def\r{\rho}
\def\s{\sigma}
\def\bs{\mbox{\boldmath $\sigma$}}
\def\t{\theta}
\def\om{\omega}


\begin{center}
{\LARGE{\bf{ SUSY approach to Pauli Hamiltonians with an axial symmetry}}}
\end{center}

\bigskip\bigskip

\begin{center}
M V Ioffe\footnote[1]{On leave of absence from Department of Theoretical Physics, Sankt-Petersburg State University, 198504 Sankt-Petersburg, Russia.}, \c{S} Kuru\footnote[2]{On leave of absence from Department of Physics, Faculty of Sciences, Ankara University
06100 Ankara, Turkey}, J Negro and L M Nieto
\end{center}

\begin{center}
{ Departamento de F\'{\i}sica Te\'{o}rica, At\'omica y \'Optica, 
Universidad de Valladolid, 47071 Valladolid, Spain}%
 \end{center}

\begin{abstract}
A two-dimensional Pauli Hamiltonian describing the interaction of a neutral spin-$1/2$
particle with a magnetic field having axial and second order symmetries, is considered. After separation of variables, the one-dimensional matrix Hamiltonian is analyzed from the point of view of supersymmetric quantum mechanics. Attention is paid to the discrete symmetries of the Hamiltonian and also to the Hamiltonian hierarchies generated by intertwining operators. 
The spectrum is studied by means of the associated matrix shape-invariance.
The relation between the intertwining operators and the second order symmetries is established and the full set of ladder operators that complete the dynamical algebra is constructed.
\end{abstract}

\noindent{PACS: 11.30.Pb, 03.65.Ge, 03.65.Fd, 02.30.Gp}


\section{Introduction}

In this work we will study a Pauli Hamiltonian describing the 
interaction of a neutral spin-$1/2$ particle interacting with
a magnetic field generated by an electric current-carrying straight wire. This system was introduced in Ref.~\cite{Pronko}, where it was analyzed in the momentum space. Here, we will carry out a systematic study in the configuration space based on the techniques of supersymmetric  (SUSY)  quantum mechanics \cite{Junker, Bagchi}, sometimes referred to as the factorization method \cite{Infeld}, a technique that has been already used to study spin-$1/2$ Pauli equations (for different approaches, see for example 
\cite{Rittenberg}).

To motivate the specific form of the Pauli Hamiltonian for the present work,  we will start with the formulation of its  symmetry properties. We will see later that this additional information is also closely related to the factorization method. Thus, let us consider the Pauli Hamiltonian in the three-dimensional space
\be\label{haminitial}
{\cal H}_3 = \frac{{\bf p}^2}{2m} + \mu\, {\bf \sigma} \cdot {\bf B}({\bf x}) + V({\bf x})
\ee
where  ${\bf B}({\bf x})$ is a magnetic field, $\mu$ is the magnetic moment of the particle, $V({\bf x})$ is a scalar potential, and ${\bf \sigma} =(\s_1,\s_2,\s_3)$, $\s_j$ being the Pauli matrices, that is,
\be\label{paulimatrices}
\s_0=\left(\begin{array}{cc} 1 & 0\\ 0& 1\end{array}\right)\quad
\s_1=\left(\begin{array}{cc} 0 & 1\\ 1&0 \end{array}\right)\quad
\s_2=\left(\begin{array}{cc} 0 & - i\\ i &0 \end{array}\right)\quad
\s_3=\left(\begin{array}{cc} 1 & 0\\ 0&-1 \end{array}\right).
\ee
We take the usual notation for the cartesian coordinates ${\bf x}=(x_1,x_2,x_3)$ and the momentum operators $p_k = -i\, \hbar\, \p/\p x_k\equiv \p x_k$. Now, we will impose some symmetries on this Hamiltonian in order to obtain the precise example we want to consider in this work.

\subsection{Rotational-translational symmetry along the $x_3$-axis}

We assume that the current-carrying wire is placed on the $x_3$-axis, and
we look for the systems (\ref{haminitial}) allowing for first order symmetries of the form
\be\label{pj3}
{\cal P}_3= p_3+ a({\bf x})
\quad
{\cal J}_3= x_1\, p_2- x_2\, p_1 + b({\bf x})\equiv 
L_3+ b({\bf x})
\ee
where $a({\bf x})$ and $b({\bf x})$ are Hermitian matrix-valued functions to be determined. By imposing that ${\cal J}_3$ and ${\cal P}_3$  are symmetries of ${\cal H}_3$, that is, $[{\cal J}_3,{\cal H}_3]=[{\cal P}_3,{\cal H}_3]=0$, we are led to the following explicit expressions (up to an equivalence):
\bea
&& {\cal J}_3 = -i\hbar\,\p_{\t} + \b\, \s_3 
	\qquad {\cal P}_3= -i\hbar\,\p_{x_3}\\
&& {\bf B}({\bf x}) = 
(f(\r) \cos [2\b(\t-\t_0)], f(\r) \sin [2\b(\t-\t_0)], g(\r))\\
&&V({\bf x}) = V(\r)
\eea
where $(\r,\t,x_3)$ are cylindrical coordinates, the functions $f(\r),g(\r),V(\r)$ are arbitrary, and $\b,\t_0$ are free parameters. Therefore, we can decouple the problem as follows: a free motion in the $x_3$--axis and another motion in the plane $(x_1,x_2)$. In the sequel, we will restrict ourselves to the analysis of the system in the plane, with Hamiltonian ${\cal H}_2={\cal H}_3- p^2_3/(2m)$.

\subsection{Parabolic symmetry}
Next, we will also assume that there exists a second order symmetry of parabolic type, that is, with a leading second order term associated to parabolic coordinates \cite{Miller},
\be
{\cal S}_1= L_3\, p_1 + p_1\, L_3 + {\bf A}({\bf x})\cdot {\bf p} +
{\bf p}\cdot {\bf A}({\bf x}) + N({\bf x})
\ee
where ${\bf A}({\bf x})$ and $N({\bf x})$ are matrix-valued vector and scalar Hermitian functions, respectively. If the magnetic term determined by $f(\r)$ is present, we arrive, up to an equivalence, to the same model already reported in \cite{Pronko}:
\bea
&&\label{b}
{\bf B}({\bf x}) = 
 ({x_2}/{\r^2}, {-x_1}/{\r^2}, 0) \qquad V({\bf x}) = 0\\[1ex]
&&\label{j3}
{\cal J}_3 = -i\hbar\, \p_{\t}  +\frac\hbar2\, \s_3
\equiv L_3 +\frac\hbar2\, \s_3\\[1ex]
&&\label{se1}
{\cal S}_1= {\cal J}_3\, p_1 + p_1\, {\cal J}_3  -\mu\, x_2\, \bs \cdot {\bf B}({\bf x}) .
\eea
The commutator of ${\cal J}_3$ and ${\cal S}_1$ gives another second order symmetry 
${\cal S}_2$:
\be\label{se2}
{\cal S}_2= {\cal J}_3\, p_2 + p_2\, {\cal J}_3  +\mu\, x_1 \, \bs \cdot {\bf B}({\bf x}) .
\ee 
All these symmetries, together with the Hamiltonian ${\cal H}_2$, close the following quadratic algebra:
\be
\label{commutjs}
[{\cal J}_3,{\cal S}_1] = i\, {\cal S}_2\quad
[{\cal J}_3,{\cal S}_2] = - i\, {\cal S}_1\quad
[{\cal S}_1,{\cal S}_2] = -4\, i\, {\cal H}_2 \, {\cal J}_3 .
\ee
As it was mentioned  in \cite{Pronko},  the symmetries ${\cal S}_1,{\cal S}_2$ are similar to the components of the Laplace-Runge-Lenz vector for the Coulomb potential.

Previous works have considered the quantum mechanical problem that we are dealing in this paper, mainly in the momentum space \cite{Voronin}, but also paying attention to some partial aspects  in the configuration space \cite{Blumel}. In the present work we want to address it from a self-contained, systematic and more complete point of view, including important properties not considered before. Basically, we will use the SUSY quantum mechanics approach with special emphasis on the shape-invariance of the model.

This paper is organized as follows. In section 2 we will perform the separation of variables and we will obtain the discrete symmetries that are basic in the development of the following sections. Next, in section 3 we analyze the factorization and the shape-invariance  properties of the radial equation. We study the ground states by means of this factorization in section 4, and the excited states in section 5.  In section 6, we will study the relationship between the second-order symmetries and the intertwining operators entering the factorization. In section 7 we build the matrix ladder operators connecting eigenstates with different energies of the same Hamiltonian. Finally, in section 8 we conclude this work with some remarks, stressing the most original results obtained in this paper.


\section{Separation of variables and discrete symmetries}\label{separation}

Let us consider again the two-dimensional matrix Hamiltonian ${\cal H}_2$ obtained from (\ref{haminitial}), where we have excluded the free part along the $x_3$-axis, and where the interaction is given by the magnetic field (\ref{b}), thus, sharing the symmetries 
(\ref{j3})--(\ref{commutjs}). In order to simplify the expressions, we take
\be
\label{hbar=1}
\hbar= {2m}=1 \quad x=\mu\, x_1\quad y=\mu\, x_2 \quad r=\mu\,\r
\ee 
so that we will work with the following form of the Hamiltonian
\be
\label{2dmodel}
{\cal H} ={\cal H}_2/\mu^2= -(\p_x^2 + \p_y^2) + \frac{y}{r^2}\, \s_1 - \frac{x}{r^2}\,\s_2 
\ee
where the second order symmetries are 
\be
\label{secondordsym}
{\cal T}_j = \mu^{-1} {\cal S}_j \quad j=1,2.
\ee

\subsection{Separation of variables}

Since this Hamiltonian commutes with the operator ${\cal J}_3$ given in (\ref{j3}),
we can look for their common eigenfunctions
\be\label{eigen}
{\cal H}\,\Psi_{\l} = E\,\Psi_{\l}\qquad
{\cal J}_3\,\Psi_{\l} = \l\, \Psi_{\l}\qquad 
\Psi_{\l}= 
\left(\begin{array}{c}\psi_1\\ \psi_2\end{array}\right) .
\ee
The rotational symmetry can be used to separate variables in polar coordinates $(r,\theta)$ so that the ${\cal J}_3$-eigenfunctions in (\ref{eigen}) are given by
\be\label{23}
\Psi_{\l}(r,\theta)= Y_{\l}(\theta) F_{\l}(r)
\ee
with
\be\label{23a}
 Y_{\l}(\theta) =\left(\begin{array}{cc} e^{i(\l-1/2)\theta} &0 \\ 
0 & e^{i(\l+1/2)\theta}\end{array}\right) 
\qquad
F_{\l}(r)=\left(\begin{array}{c}f_1(r)\\ f_2(r)\end{array}\right).
\ee
Now, taking into account the polar expression for the Laplacian 
$\Delta = \p_r^2 +1/r\, \p_r + 1/r^2\, \p_{\theta}^2$, the eigenfunction equation for $\cal H$ takes the form
\be
-F_{\l}''(r)-\frac1r\, F_{\l}'(r) +\frac{\l^2+\frac14}{r^2} F_{\l}(r) - 
\frac{\l}{r^2}\, \s_3 \, F_{\l}(r)
-\frac{1}{r}\, \s_2 \,  F_{\l}(r) = E \, F_{\l}(r).
\ee
In order to eliminate the first order derivative term, we make the replacement
\be\label{ff}
F_{\l}(r) = r^{-1/2} \Phi_{\l}(r)
\ee
so that we finally have a one-dimensional matrix Schr\"odinger-like equation
\be\label{pauli}
{\cal H}_{\l}\Phi_{\l}(r) 
\equiv
\left\{
-\frac{d^2}{d r^2}+\frac{\l^2 - \l\, \s_3}{r^2} 
- \frac{\s_2}{r} \right\}\Phi_{\l}(r)  =  E\,\Phi_{\l}(r) 
\ee
where we have introduced the notation ${\cal H}_{\l}$ for the radial part of the initial Hamiltonian ${\cal H}$.
Notice that this equation is quite similar to the radial Schr\"odinger equation for a charged particle in a Coulomb potential
\be\label{coulomb}
{\cal H}_{\ell}^c\phi(r) \equiv
-\phi''(r)+ \frac{\ell^2 - \ell}{r^2}\,\phi(r) - \frac{1}{r}\,\phi(r) =  E^c\,\phi(r)
\ee
where $\ell$ is the orbital momentum. The discrete spectrum is given by the well known formula
\be\label{cspectrum}
E^c = -\frac1{4(\ell +n+1)^2} \qquad n=0,1,\dots
\ee


\subsection{Discrete symmetries} \label{discrete} 
In the following we will describe some discrete symmetries of the matrix equation (\ref{pauli}).

\subsubsection{Conjugation.}
Let us consider the antilinear operator
\be
{\cal C}= \s_3 \, {\cal K}
\ee
where ${\cal K}$ is the complex conjugation operator, ${\cal K}\,\Phi(r) = \Phi^*(r)$. Then, it is immediate to check that this is a symmetry of ${\cal H}_{\l}$:
\be
{\cal H}_{\l}\, {\cal C} = {\cal C}\, {\cal H}_{\l} .
\ee
The eigenfunctions of ${\cal C}$, up to a global phase factor, are of the form:
\be\label{conjugation}
\Phi= 
\left(\begin{array}{c}\phi_1(r)\\ i\, \phi_2(r)\end{array}\right)
\ee
where $\phi_1(r)$ and $\phi_2(r)$ are real functions. Hence, from now on, we will choose the eigenfunctions of the matrix equation (\ref{pauli}) in the form (\ref{conjugation}).

\subsubsection{Reflection in $\l$.}
It is very easy to see from (\ref{pauli}) that
\be\label{s22}
\s_2\, {\cal H}_{\l} ={\cal H}_{-\l}\, \s_2 .
\ee
This means that the eigenfunction problem for ${\cal H}_{\l}$ is equivalent to that of 
${\cal H}_{-\l}$:
\be
{\cal H}_{\l} \Phi_{\l} = E\, \Phi_{\l} \quad  \Longleftrightarrow \quad 
{\cal H}_{-\l} \Phi_{-\l} = E\, \Phi_{-\l} \quad
\Phi_{-\l} = \s_2\, \Phi_{\l} .
\ee
Observe that these two discrete symmetries can be implemented also in the space of eigenfunctions of the two-dimensional equations (\ref{eigen}).
If we want to compare these discrete symmetries with the ones of the Coulomb Hamiltonian ${\cal H}^c_{\ell}$ of (\ref{coulomb}), we see that the conjugation is translated into the real character of the eigenfunctions, while the reflection property means that  ${\cal H}^c_{\ell} = {\cal H}^c_{-\ell+1}$.


\section{Factorizations and shape-invariance}\label{factor}

In this section we will investigate the factorization and
the supersymmetrical properties of (\ref{pauli}), the radial part of Pauli
equation. This one-dimensional $2\times2$ matrix problem is 
interesting by itself, for an arbitrary value of $\lambda$, but especially  for the 
two-dimensional physical model (\ref{2dmodel}) with $\lambda=1/2+m, \, m\in\Z$. 
In particular, we will see that this matrix model obeys the shape-invariance properties. This is very interesting because, up to now, the shape-invariance was used as a very elegant algebraic approach in the solution of one and two-dimensional scalar
spectral problems \cite{Junker,Bagchi,Gendenstein,Cannata}, while in this section it will be used to determine the spectrum of a one-dimensional matrix example.

Following closely the well known factorization of the Coulomb Hamiltonian (\ref{coulomb}), we propose here the following ansatz for this matrix case:
\be
\label{not1}
{\cal H}_{\l} = \left(\frac{d}{d r} +\frac{A+B\, \s_3}{r}+ D\, \s_2 \right)
\left(-\frac{d}{d r} +\frac{A+B\, \s_3}{r}+ D\, \s_2 \right) + \g 
\ee
where $A,B,D$ and $\g$ are real constants to be determined. Indeed, we find two different solutions having the following form
\be\label{not}
{\cal H}_{\l} = L^-_{\l}L^+_{\l}+ \g_{\l}  =L^+_{\l-1}L^-_{\l-1}+ \g_{\l-1}
\qquad
\g_{\l}  = \frac{-1}{4(\l+1/2)^2}
\ee
where we have used the following notation for the operators
\be
\label{not3}
L^\pm_{\l}=
\mp\frac{d}{d r} +\frac{(\l+1/2) -(1/2) \s_3}{r} - 
\frac{(1/2) \s_2}{(\l+1/2)}  \equiv 
\mp\frac{d}{d r} + W_{\l}(r)
\ee
being $W_{\l}(r)$ the matrix superpotentials.

Of course, equations  (\ref{not1})--(\ref{not3}) are valid for any value of $\l\neq -1/2$, and therefore we can build a hierarchy of Hamiltonians $\{{\cal H}_{\l+m}\}$, $m\in \Z$, as follows:
\be\label{recurrence}
{\cal H}_{\l+m} = L^-_{\l+m}L^+_{\l+m}+ \g_{\l+m}=L^+_{\l+m-1}L^-_{\l+m-1}+ 
\g_{\l+m-1} .
\ee
These relations allow to define an algebra of operators of the Hamiltonian hierarchy 
\cite{olmo}. 
The operators $L^{\pm}_{\l+m}$ act as intertwining operators between two consecutive Hamiltonians ${\cal H}_{\l+m}$ and ${\cal H}_{\l+m+1}$:
\be\label{intert}
L^+_{\l+m}{\cal H}_{\l+m} ={\cal H}_{\l+m+1}L^+_{\l+m}\quad  
{\cal H}_{\l+m}L^-_{\l+m} =L^-_{\l+m} {\cal H}_{\l+m+1} .
\ee
These relations imply that the eigenfunctions of ${\cal H}_{\l+m}$ can be obtained from those of ${\cal H}_{\l+m+1}$ with the same eigenvalue by acting with $L^-_{\l+m}$, while the operator $L^+_{\l+m}$ makes this connection in the opposite way. This correspondence between square-integrable eigenfunctions is one-to-one, except when $L^{\pm}_{\l+m}$ annihilates some eigenfunctions. As usual, relations (\ref{recurrence}) and (\ref{intert}) mean that the Hamiltonians of the hierarchy $\{{\cal H}_{\l+m}\}$ are shape-invariant: intertwined Hamiltonians have the same shape but different values of the parameter.

Now, let us see the effect of the discrete symmetries ${\cal C}$ and $\s_2$ on the intertwining operators. It is easy to check that
\be\label{c}
{\cal C}\, L^{\pm}_{\l+m}= L^{\pm}_{\l+m}\, {\cal C}
\ee 
and
\be\label{ss2}
\s_2 \,L^+_{\l +m} = L^-_{-\l -m-1}\, \s_2 .
\ee
From the commutation (\ref{c}) we see that the interwining operators keep the form of the eigenfunctions $\Phi$ in (\ref{conjugation}). Relation (\ref{ss2}) will have consequences for the ground states, as we will see in the next section.

The hierarchies obtained from values $\l$ and $\l+m_0$, where 
$m_0\in\Z,$ are the same, so we can get a family of different hierarchies characterised, for instance, by the values $\l\in [0,1/2]$. 
Although the initial two-dimensional model (\ref{2dmodel}) is related to the hierarchy with half-integer values of $\l$, we will analyze the properties for any $\l$. Thus, we can distinguish the following three classes of hierarchies.
\begin{itemize}
\item 
$\l=0$. 
We call it the {\it `integer hierarchy'} $\{{\cal H}_{m}\}$, $m\in \Z$. It includes the Hamiltonian ${\cal H}_0$ and would correspond to two-valued eigenfunctions in the context of the two-dimensional problem  (\ref{2dmodel}). The reflection in $\l$ is implemented  in the hierarchy, a scheme of which can be seen in Figure~\ref{fig1}.

\item
$\l=1/2$. 
We refer to this case as the {\it `physical hierarchy'} $\{{\cal H}_{m+1/2}\}$, $m\in\Z$, because it is associated to the physical spin-$1/2$ systems of Eq.~ (\ref{2dmodel}). Indeed, it is the only case considered in all the previous references 
\cite{Pronko,Voronin,Blumel}, and for it the operators 
$L^{\pm}_{-1/2}$ are not well defined, as can be seen from (\ref{not3}).  Therefore, there are no first-order intertwining operators connecting the Hamiltonians ${\cal H}_{1/2}$ and ${\cal H}_{-1/2}$. Remark that the hierarchy also includes the reflection in $\l$.
\item
$0<\l<1/2$. 
We will refer to this as the {\it `general hierarchy'} $\{{\cal H}_{\l+m}\}$. It does not implement the reflection in $\l$; in fact this reflection gives rise to the hierarchies with values $-1/2<\l<0$. The associated two-dimensional eigenfunctions are multiple-valued.
\end{itemize}


\section{The ground states of the hierarchies}

It is known \cite{Junker,Gendenstein} that the ground states of SUSY-hierarchy
Hamiltonians are the main elements when using the shape-invariance
approach in the construction of the whole spectra and the eigenfunctions
for one-dimensional scalar problems. In this section, the explicit expressions for the ground states of the matrix SUSY-hierachies described in the previous section will be found and analyzed.

Let us start with the ground-state wave-functions $\Phi_{\l+m}^0$ annihilated by the general intertwining operator $L^+_{\l+m}$:
\be\label{ground}
L^+_{\l+m}\Phi_{\l+m}^0=0 .
\ee
Then, according to (\ref{not}) and  (\ref{recurrence}), this will be an eigenfunction of ${\cal H}_{\l+m} $ with
energy $E_{\l+m}^0=\gamma_{\l+m}$:
\be
{\cal H}_{\l+m} \Phi_{\l+m}^0= E_{\l+m}^0 \Phi_{\l+m}^0 \qquad
E_{\l+m}^0= -\frac{1}{4(\l +m+1/2)^2}.
\ee
Let us use the notation (\ref{conjugation}) for the ground state 
\be
\Phi_{\l+m}^0= \left(\begin{array}{c}\phi_{1,\l+m}^0(r)\\[1.ex]
i\, \phi_{2,\l+m}^0(r) \end{array}\right)
\ee
and let us make the substitution
\be
\phi_{j,\l}^0(r) = z^{1+\l+m}\varphi_j(z)\quad j=1,2 \qquad 
z = \frac{r}{2(\l+m+1/2)}
\ee
in equation (\ref{ground}). Then, we arrive to a modified Bessel equation for the component $\varphi_1(z)$
\be
z^2 \varphi_1''(z) + z\, \varphi_1'(z) - 
\left(z^2 +1\right) \varphi_1(z)=0
\ee 
with general solution
\be
\varphi_1(z) = c_1\, I_1(z) +c_2\, K_1(z)
\ee
where $I_1(z), K_1(z)$ are the two linearly independent modified Bessel functions. After a simple calculation, we arrive to the general form of the ground state (\ref{ground}) for any integer $m$:
\be
\label{KIlm}
\Phi_{\l+m}^0(r)= c_1\,{\mathtt K}_{\l+m}(r)+ c_2\,{\mathtt I}_{\l+m}(r)
\ee
with
\bea
\label{sol1}
&&
{\mathtt K}_{\l+m}(r) =   r^{1+\l+m}
\left(\begin{array}{c}K_1  \left(\frac{r}{2(\l+m+1/2)}\right) \\[1.ex]
i\, K_0 \left(\frac{r}{2(\l+m+1/2)}\right) \end{array}\right) 
\\ [1ex]
&&
{\mathtt I}_{\l+m}(r)= r^{1+\l+m}
\left(\begin{array}{c}I_1 \left(\frac{r}{2(\l+m+1/2)}\right) \\[1.ex]
-i\, I_0 \left(\frac{r}{2(\l+m+1/2)} \right)\end{array}\right) .
\label{sol2}
\eea

Remark that as the asymptotic behaviour of the modified Bessel functions  $I_0(r)$ and $I_1(r)$ in the limit $r\to +\infty$ is exponentially increasing, they can not lead to square-integrable functions. On the other side, in the same limit $K_0(r)$ and $K_1(r)$ decrease exponentially, while near the origin $K_0(r) \approx \log(r)$ and $ K_1(r) \approx 1/{r}$. Thus, the physical solution will be $\Phi_{\l+m}^0(r)\propto {\mathtt K}_{\l+m}(r)$, which is a well behaved ground state of ${\cal H}_{\l+m}$ with $\l+m> 0$ (the ground state of ${\cal H}_{0}$ will be considered later).

Let us consider now the ground states $\widetilde \Phi_{\l+m}^0$ of ${\cal H}_{\l+m}$ with $\l+m< 0$. They are annihilated by the operator $L^-_{\l+m-1}$ in the factorization (\ref{not}):
$L^-_{\l+m-1}\widetilde\Phi_{\l+m}^0=0$.
Then, according to  (\ref{not}) and  (\ref{recurrence}), this will be an eigenfunction with
energy $\widetilde E_{\l+m}^0 =E_{-\l-m}^0=\gamma_{-\l-m}$:
\be
{\cal H}_{\l+m} \w \Phi_{\l+m}^0= \w E_{\l+m}^0 \w \Phi_{\l+m}^0 \qquad
\w E_{\l+m}^0= - \frac{1}{4(\l +m-1/2)^2} .
\ee
After a computation similar to the one carried out in the previous case, we get the following general solutions:
\bea
\w\Phi_{\l+m}^0(r)&=& c_1\, r^{1-\l-m}
\left(\begin{array}{c}
K_0 \left( \frac{-r}{2(\l+m-1/2)} \right)\\[1.5ex]
i\, K_1 \left(\frac{-r}{2(\l+m-1/2)} \right)\end{array}\right) \nonumber \\
&& +c_2\, r^{1-\l-m}
\left(\begin{array}{c}
I_0 \left(\frac{-r}{2(\l+m-1/2)} \right)\\[1.5ex]
-i\, I_1 \left(\frac{-r}{2(\l+m-1/2)} \right)\end{array}\right) .
\label{secondtilde}
\eea
As in the previous case, a study of the asymptotic behaviour shows that the square-integrable and finite ground states are obtained by taking $c_2=0$ in (\ref{secondtilde}),  whenever $\l+m<0$. The results for this case could be found directly from the previous one with the help of the relation (\ref{ss2}) between both types of intertwining operators through $\s_2$.

In summary, we have one reasonable ground state for each Hamiltonian 
${\cal H}_{\l+m}$: if $\l+m> 0$ it is of the form $\Phi_{\l+m}^0(r)$, while if $\l+m< 0$ it will take the form $\w\Phi_{\l+m}^0(r)$. There is only one Hamiltonian, ${\cal H}_0$, having a doubly degenerated square-integrable solution with energy $E_{0}^0= - 1$. But this case is rather special and we will comment on its lack of physical meaning later on.

To end this section, let us discuss now the relationship between these ground-state solutions and the superpotential. First of all, let us recall that
\be\label{ki}
L^+_{\l+m}\,{\mathtt K}_{\l+m}(r)=L^+_{\l+m}\,{\mathtt I}_{\l+m}(r)=0.
\ee
Let us introduce now the $2\times 2$ matrix ${\mathbb M}_{\l+m}(r)$, whose columns are 
${\mathtt K}_{\l+m}(r)$ and  ${\mathtt I}_{\l+m}(r)$, in this order. Then, equations (\ref{ki}) can be expressed in a more compact form:
\be
L^+_{\l+m}\,{\mathbb M}_{\l+m}(r)=-{\mathbb M}_{\l+m}'(r)+ W_{\l+m}(r)\,{\mathbb M}_{\l+m}(r)=0 .
\ee 
Thus, after substituting (\ref{KIlm}), we get again the expression of (\ref{not3}) for the matrix superpotential in terms of the solution matrix:
\be\label{w}
W_{\l+m}(r)={\mathbb M}_{\l+m}'(r) \, {\mathbb M}^{-1}_{\l+m}(r)=
\frac{(\l+m+\frac12) -\frac12 \s_3}{r} - 
\frac{ \frac12 \s_2}{\l+m+\frac12}  .
\ee
Therefore, we have shown that this class of matrix solutions ${\mathbb M}_{\l+m}(r)$ gives rise in a nontrivial way to a Hermitian matrix superpotential. It is important to stress that there are not many explicit examples satisfying this condition 
\cite{boris}.


\section{Excited states}\label{excited}

In this section we will study the excited states  of the system (\ref{pauli}). There are some general expressions valid for any hierarchy.
In order to find the $n$-excited state $\Phi_{\l}^n(r)$ of a particular Hamiltonian  ${\cal H}_{\l}$, $\l>0$, we can start with the ground state $\Phi_{\l+n}^0(r)$, $n\in\Z^+$, of the Hamiltonian ${\cal H}_{\l+n}$ in the same hierarchy. Then
\be\label{exc}
\Phi_{\l}^n(r)=L_{\l}^-L_{\l+1}^-\cdots L_{\l+n-1}^-\, \Phi_{\l+n}^0(r)
\ee
with energy
\be
\label{excenergy}
E_{\l}^n= -\frac{1}{4(\l + n + 1/2)^2} .
\ee

Two natural questions arise now: 
\begin{itemize}
\item[a)] 
Are the excited states obtained in this way always square-integrable? 
\item[b)] 
Are these excited states the only bounded physical states for each Hamiltonian of the hierarchy? 
\end{itemize}
Here, in the context of the factorization method, we can answer to the first question: under the assumption $\l>0$, all the eigenfunctions build in the form 
(\ref{exc}) are finite, square-integrable and vanishing at the endpoints ($0$ and $+\infty$), so they indeed describe excited bound states. The details are given in the Appendix. 
With respect to the second question, as in the two-dimensional scalar models 
\cite{Cannata}, it should be studied from the point of view of a general ``oscillation theorem" for matrix Sturm-Liouville operators to guarantee that no additional excited states exists besides those obtained after applying the shape-invariance procedure (for a discussion on two-dimensional scalar models see \cite{Cannata}). Meanwhile, one can notice that each component of the constructed $n$-th excited states (\ref{exc}) has exactly $(n-1)$ nodes.

If $\l <0$, the excited states with similar properties are obtained from the other class of ground states:
\be\label{exc2}
\w\Phi_{\l}^n(r)=L_{\l-1}^+L_{\l-2}^+\cdots L_{\l-n}^+\, \w\Phi_{\l-n}^0(r)
\ee
with energy
\be
\w E_{\l}^n= -\frac{1}{4(\l - n - 1/2)^2} .
\label{exc44}
\ee
Now, we will consider some specific features of each hierarchy.

\subsection
{`Integer hierarchy' ($\l=0$): ${\cal H}_{m}$, $m\in \Z$.}

A representation of this hierarchy can be seen in Figure~\ref{fig1}. 
It includes the Hamiltonian ${\cal H}_{0}$ which has the peculiarity that both procedures lead to two square-integrable excited states for each energy
\be
E_{0}^n= -\frac{1}{4( n + 1/2)^2} .
\ee
Therefore, in principle, each eigenvalue level is doubly degenerated. However, an important difference is that in each eigenfunction one of the components vanish at the origin, while the other one does not. This feature leads to some problems about the physical interpretation, that we will discuss now.

\begin{center}
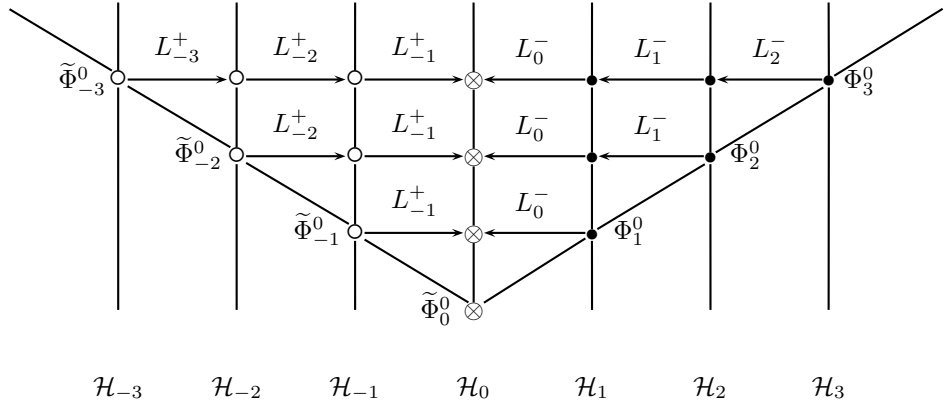
\begin{figure}[htp]
\begin{picture}(350,160)
\put(-5,-20)
{    \begin{psmatrix}[rowsep=0.6cm,colsep=0.9cm]
\phantom{\Large$\circ$}&{} & {} & {} & {} & {} & {} & {} & \phantom{ } \\
{}& {\Large$\circ$} & {\Large$\circ$} & {\Large$\circ$} &{$\otimes$} & $\bullet$ & $\bullet$ &$\bullet$ 
\\
{}&{} & {\Large$\circ$} & {\Large$\circ$} & $\otimes$ & $\bullet$ & $\bullet$ & {}\\
{}&{} & {} & {\Large$\circ$} &$\otimes$ & $\bullet$ & {} & {}\\
{}&{} & {} & {} & $\otimes$ & {} & {} & {}\\
{} & ${\cal H}_{-3}$ & ${\cal H}_{-2}$ &${\cal H}_{-1}$ & ${\cal H}_{0}$ & ${\cal H}_{1}$ & ${\cal H}_{2}$ &${\cal H}_{3}$ & {} \\
 \phantom{${\cal H}_{-3}$}& \phantom{${\cal H}_{-3}$} &  \phantom{${\cal H}_{-3}$} &  \phantom{${\cal H}_{-3}$} &  \phantom{${\cal H}_{-3}$} &  \phantom{${\cal H}_{-3}$} &  \phantom{${\cal H}_{-3}$} &  \phantom{${\cal H}_{-3}$} &  \phantom{${\cal H}_{-3}$}
 %
      \ncline{->}{2,2}{2,3}^{$L_{-3}^+$}
      \ncline{->}{2,3}{2,4}^{$L_{-2}^+$}
      \ncline{->}{2,4}{2,5}^{$L_{-1}^+$}
      \ncline{<-}{2,5}{2,6}^{$L_{0}^-$}
      \ncline{<-}{2,6}{2,7}^{$L_{1}^-$}
      \ncline{<-}{2,7}{2,8}^{$L_{2}^-$}
      \ncline{->}{3,3}{3,4}^{$L_{-2}^+$}
      \ncline{->}{3,4}{3,5}^{$L_{-1}^+$}
      \ncline{<-}{3,5}{3,6}^{$L_{0}^-$}
      \ncline{<-}{3,6}{3,7}^{$L_{1}^-$}
      \ncline{->}{4,4}{4,5}^{$L_{-1}^+$}
      \ncline{<-}{4,5}{4,6}^{$L_{0}^-$}
 \ncdiag[armA=0,armB=0,angleA=35,angleB=-125]{-}{2,8}{1,9}
 \ncdiag[armA=0,armB=0,angleA=35,angleB=-125]{-}{3,7}{2,8}
 \ncdiag[armA=0,armB=0,angleA=35,angleB=-125]{-}{4,6}{3,7}
 \ncdiag[armA=0,armB=0,angleA=35,angleB=-140]{-}{5,5}{4,6}
 \ncdiag[armA=0,armB=0,angleA=125,angleB=-35]{-}{2,2}{1,1}
 \ncdiag[armA=0,armB=0,angleA=125,angleB=-35]{-}{3,3}{2,2}
 \ncdiag[armA=0,armB=0,angleA=125,angleB=-35]{-}{4,4}{3,3}
 \ncdiag[armA=0,armB=0,angleA=125,angleB=-35]{-}{5,5}{4,4}
      \ncline{-}{5,2}{2,2}
      \ncline{-}{2,2}{1,2}
      \ncline{-}{5,3}{3,3}
      \ncline{-}{3,3}{2,3}
      \ncline{-}{2,3}{1,3}
      \ncline{-}{5,4}{4,4}
      \ncline{-}{4,4}{3,4}
      \ncline{-}{3,4}{2,4}
      \ncline{-}{2,4}{1,4}
	\ncline{-}{5,5}{4,5}
	\ncline{-}{4,5}{3,5}
	\ncline{-}{3,5}{2,5}
	\ncline{-}{2,5}{1,5}
      \ncline{-}{5,6}{4,6}
      \ncline{-}{4,6}{3,6}
      \ncline{-}{3,6}{2,6}
      \ncline{-}{2,6}{1,6}
      \ncline{-}{5,7}{3,7}
      \ncline{-}{3,7}{2,7}
      \ncline{-}{2,7}{1,7}
      \ncline{-}{5,8}{2,8}
      \ncline{-}{2,8}{1,8}
    \end{psmatrix}
}
\put(30,122){$\widetilde{\Phi}_{-3}^0$}
\put(74,94){$\widetilde{\Phi}_{-2}^0$}
\put(119,65){$\widetilde{\Phi}_{-1}^0$}
\put(167,35){$\widetilde{\Phi}_{0}^0$}
\put(240,65){${\Phi}_{1}^0$}
\put(284,94){${\Phi}_{2}^0$}
\put(327,122){${\Phi}_{3}^0$}
\end{picture}
\caption{Energy level diagram for the `integer hierarchy', including the information about the ground states and the operators $L^\pm_k$.} \label{fig1}
\end{figure}
\end{center}

\begin{figure}
  \centering
   \fbox{
\includegraphics[width=0.4\textwidth]{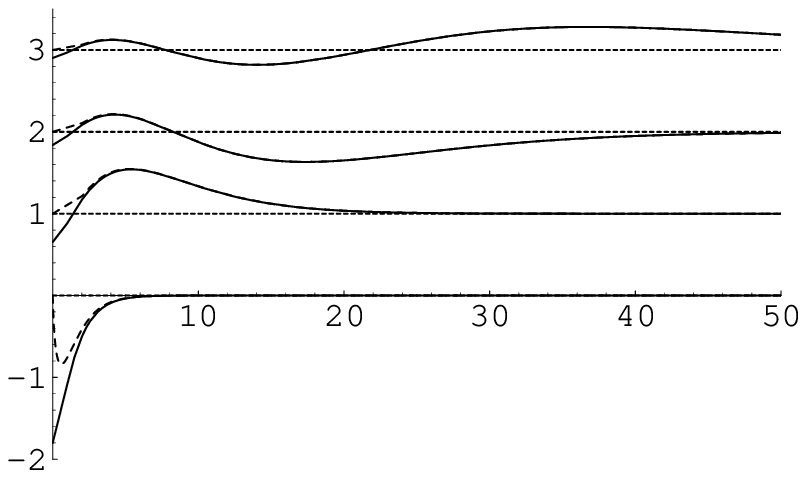}
        }%
\hspace{1cm}%
   \fbox{
  \includegraphics[width=0.4\textwidth]{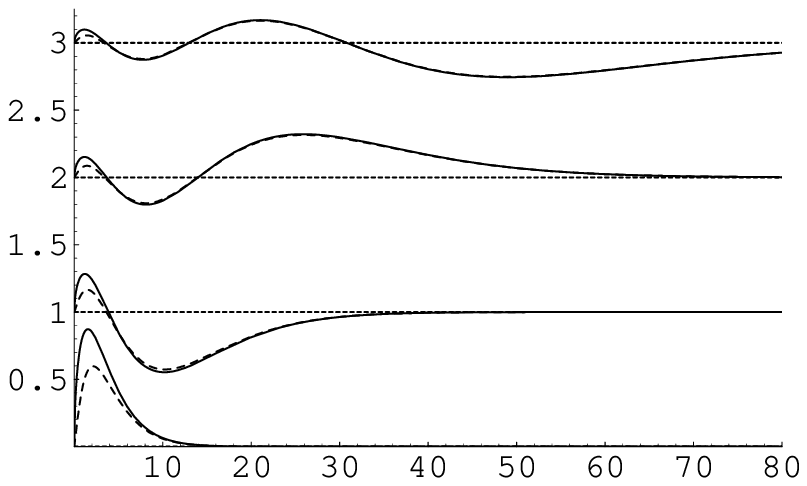}
     }
\caption{Plot of the two components of the eigenfunctions (the first component in solid line, the second one in dashed line) for the ground and the first excited states of the Hamiltonians ${\cal H}_0$ (left) and ${\cal H}_{1/2}$ (right).}
  \label{dosfiguras}
\end{figure}

The matrix Hamiltonian ${\cal H}_0$ in (\ref{pauli}) can be diagonalized by means of the unitary matrix 
${\cal U}=\frac{1}{\sqrt2}(i \s_0+\s_1)$, 
\be
{\cal H}_0 \Phi =\left\{-\frac{d^2}{d r^2} - \frac{\s_2}{r}\right\}\Phi
\ 
\Longleftrightarrow
\ 
{\cal H}_d\Phi_d =\left\{-\frac{d^2}{d r^2} - \frac{\s_3}{r}\right\}\Phi_d
\ee
where ${\cal H}_d = {\cal U} {\cal H}_0 {\cal U}^\dag$ and $\Phi_d = {\cal U} \Phi$. Thus, we can write
\be
\label{HDD}
{\cal H}_d= \left(
\begin{array}{cc}
h_-&0\\
0&h_+
\end{array}\right) 
\ee
where  $h_{\pm} = -\frac{d^2}{d r^2} \pm \frac1r$ are scalar Hamiltonians.
In particular, the lowest energy ($E=-1$) normalizable solutions of the Schr\"odinger equation with Hamiltonian (\ref{HDD}) are as follows
\bea
&& {\Phi}_-^0 =  
\frac{r}{\sqrt2}\left(\begin{array}{c}
K_1(r) + K_0(r)\\0\end{array}\right)
\quad
{\Phi}_+^0=  
\frac{r}{\sqrt2}\left(\begin{array}{c}
0\\K_1(r) - K_0(r)\end{array}\right) 
\\
&&\Phi_d^0= {\cal U}\,\Phi_0^0\qquad 
\w \Phi_d^0= {\cal U}\,\widetilde \Phi_0^0 .
\label{sol59}
\eea
One can notice that these solutions, as well as the higher energy solutions, do not vanish at the origin, and therefore correspond to divergent mean values of the potential energy terms. A plot of some of these functions can be seen in Figure~\ref{dosfiguras}. By this reason, solutions (\ref{sol59}), though being normalizable, {\it do not\/} belong to the physical sector of the model.



Nevertheless, these non-physical solutions are quite useful from the point of view of SUSY quantum mechanics. Indeed, they are used to generate the intertwining operators
 $L_{m}^{\pm}$ to get a hierarchy ${\cal H}_m$ of shape-invariant 
non-diagonal Hamiltonians 
starting from ${\cal H}_0$, which is diagonal.  Besides, any ${\cal H}_m$, $m\neq0$ has a sensible physical discrete spectrum with eigenfunctions vanishing at the origin. By extending the label $m$ to a real number we get all the hierarchies ${\cal H}_{\l+m}$. 

\subsection
{`Physical hierarchy' ($\l=1/2$): ${\cal H}_{m+1/2}$, $m\in \Z$.}

The general scheme (\ref{exc})--(\ref{exc44}) is still valid here. But since in this hierarchy the operators $L^\pm_{-1/2}$ are not defined, in order to connect the excited states of ${\cal H}_{1/2}$ and ${\cal H}_{-1/2}$ we need a zero-order intertwining operator given by the reflection  (\ref{s22})
\be\label{sigma}
\Phi_{1/2}^n \propto \s_2\, \Phi_{-1/2}^n .
\ee

It is interesting to remark that it is possible to include in a very natural way the operator 
$\s_2$ inside the set of intertwining operators $L^{\pm}_{\l +m}$ if we `normalize' them by a factor:
\be\label{norm}
\w L^\pm_{\l +m}   := (\l+m+1/2)L^\pm_{\l +m} .
\ee
Then, the set $\{\w L^{\pm}_{\l +m}\}$, $m\in \Z$, will act also as intertwining operators of the hierarchy ${\cal H}_{\l+m}$ as in (\ref{intert}), but they are always well defined, and in particular $\w L^{\pm}_{-1/2}= -\s_2$, as it should be. 

However, the expression of the Hamiltonians in terms of the normalized operators are changed by these factors and it gives rise to the following `commutation rules':
\be\label{normalc}
\w L^+_{\l +m-1} \w L^-_{\l +m-1} -
\w L^-_{\l +m} \w L^+_{\l +m} = -2 (\l+m) {\cal H}_{\l+m}.
\ee

\subsection
{`General hierarchy' ${\cal H}_{\l+m}$, $m\in \Z$.}

If $0<\l<1/2$ we will have two kinds of Hamiltonians in the hierarchy. The right-hand Hamiltonians ${\cal H}_{\l+m}$, $m =0,1,\dots$, and the left-hand ones ${\cal H}_{\l-m}$, $m\in \N$. We have two special features: (a) The excited states of the right (left) Hamiltonians are only obtained from the right (left) ground states. However 
$L^{\pm}_{\l-1}$ do not connect these two kinds of Hamiltonians, neither the reflection matrix $\s_2$ can be used to connect the two Hamiltonian sectors. (b) The spectrum of these two types of Hamiltonians are different.  In this case the operators 
$L^{\pm}_{\l+m}$, acting on physical states, do not always generate again physical states. A schematic diagram of this case can be seen in Figure~\ref{fig3}.

\begin{center}
\begin{figure}[htp]
\begin{picture}(350,160)
\put(-10,-20)
{    \begin{psmatrix}[rowsep=0.6cm,colsep=0.8cm]
\phantom{\Large$\circ$}&{} & {} & {} & \phantom{ } \\
{}& {\Large$\circ$} & {\Large$\circ$} & {\Large$\circ$} & {\Large$\circ$}  \\
{}&{} & {\Large$\circ$} & {\Large$\circ$} & {\Large$\circ$}\\
{}&{} & {} & {\Large$\circ$} & {\Large$\circ$}\\
{}&{} & {} & {} & {\Large$\circ$}\\
{} & ${}^{{\cal H}_{\l-4}}$ & ${}^{{\cal H}_{\l-3}}$ & ${}^{{\cal H}_{\l-2}}$   & ${}^{{\cal H}_{\l-1}}$  
 & ${}^{{\cal H}_{\l}}$ & ${}^{{\cal H}_{\l+1}}$ & ${}^{{\cal H}_{\l+2}}$ & ${}^{{\cal H}_{\l+3}}$  \\
      \ncline{->}{2,2}{2,3}^{$L_{\l-4}^+$}
      \ncline{->}{2,3}{2,4}^{$L_{\l-3}^+$}
      \ncline{->}{2,4}{2,5}^{$L_{\l-2}^+$}
      \ncline{->}{3,3}{3,4}^{$L_{\l-3}^+$}
      \ncline{->}{3,4}{3,5}^{$L_{\l-2}^+$}
      \ncline{->}{4,4}{4,5}^{$L_{\l-2}^+$}
    \end{psmatrix}
 \ncdiag[armA=0,armB=0,angleA=125,angleB=-35]{-}{2,2}{1,1}
 \ncdiag[armA=0,armB=0,angleA=125,angleB=-35]{-}{3,3}{2,2}
 \ncdiag[armA=0,armB=0,angleA=125,angleB=-35]{-}{4,4}{3,3}
 \ncdiag[armA=0,armB=0,angleA=125,angleB=-35]{-}{5,5}{4,4}
      \ncline{-}{5,2}{2,2}
      \ncline{-}{2,2}{1,2}
      \ncline{-}{5,3}{3,3}
      \ncline{-}{3,3}{2,3}
      \ncline{-}{2,3}{1,3}
      \ncline{-}{5,4}{4,4}
      \ncline{-}{4,4}{3,4}
      \ncline{-}{3,4}{2,4}
      \ncline{-}{2,4}{1,4}
	\ncline{-}{5,5}{4,5}
	\ncline{-}{4,5}{3,5}
	\ncline{-}{3,5}{2,5}
	\ncline{-}{2,5}{1,5}
}
\hskip5.4cm
{    \begin{psmatrix}[rowsep=0.6cm,colsep=1.3cm]
\phantom{$\bullet$} & {} & {} & {} & \phantom{ } & {}\\
  {}& $\bullet$ & $\bullet$ & $\bullet$ & $\bullet$ &  {}  \\
 {} & $\bullet$ & $\bullet$ &$\bullet$  & {}  & {}\\
 {} & $\bullet$ & $\bullet$ & {} & {}   & {}\\
 {} & $\bullet$ & {} & {} & {}   & {}\\
      \ncline{->}{4,3}{4,2}^{$L_{\l}^-$}
      \ncline{->}{3,3}{3,2}^{$L_{\l}^-$}
      \ncline{->}{2,3}{2,2}^{$L_{\l}^-$}
      \ncline{->}{3,4}{3,3}^{$L_{\l+1}^-$}
      \ncline{->}{2,4}{2,3}^{$L_{\l+1}^-$}
      \ncline{->}{2,5}{2,4}^{$L_{\l+2}^-$}
    \end{psmatrix}
 \ncdiag[armA=0,armB=0,angleA=35,angleB=-125]{-}{2,5}{1,6}
 \ncdiag[armA=0,armB=0,angleA=35,angleB=-125]{-}{3,4}{2,5}
 \ncdiag[armA=0,armB=0,angleA=35,angleB=-125]{-}{4,3}{3,4}
 \ncdiag[armA=0,armB=0,angleA=35,angleB=-140]{-}{5,2}{4,3}
      \ncline{-}{5,5}{4,5}
      \ncline{-}{4,5}{3,5}
      \ncline{-}{3,5}{2,5}
      \ncline{-}{2,5}{1,5}
      \ncline{-}{5,4}{4,4}
      \ncline{-}{4,4}{3,4}
      \ncline{-}{3,4}{2,4}
      \ncline{-}{2,4}{1,4}
      \ncline{-}{5,3}{3,3}
      \ncline{-}{3,3}{2,3}
      \ncline{-}{2,3}{1,3}
      \ncline{-}{5,2}{2,2}
      \ncline{-}{2,2}{1,2}
\put(-160,25){${\Phi}_{\l}^0$}
\put(-118,55){${\Phi}_{\l +1}^0$}
\put(-78,83){${\Phi}_{\l +2}^0$}
\put(-34,113){${\Phi}_{\l +3}^0$}
\put(-363,125){$\widetilde{\Phi}_{\l -4}^0$}
\put(-318,95){$\widetilde{\Phi}_{\l -3}^0$}
\put(-275,66){$\widetilde{\Phi}_{\l-2}^0$}
\put(-230,37){$\widetilde{\Phi}_{\l-1}^0$}
}
\end{picture}
\caption{Energy level diagram for the `general hierarchy', including the information about the ground states and the operators $L^\pm_{\l+m}$.} 
\label{fig3}
\end{figure}
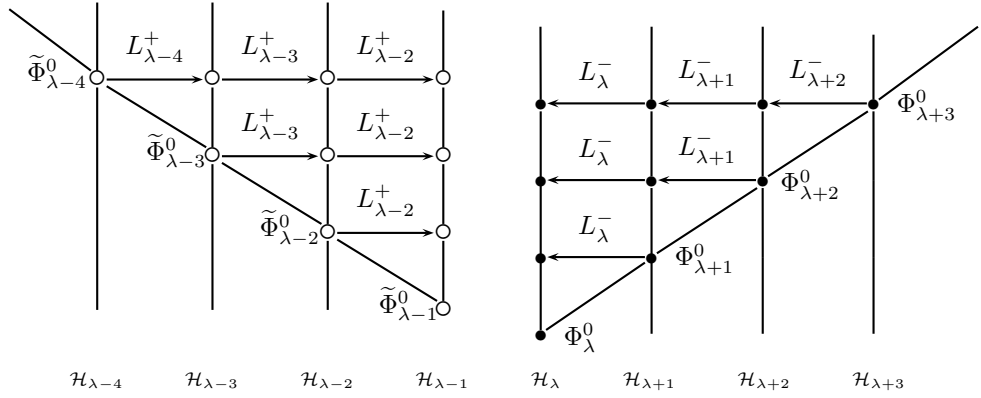
\end{center}


\section{Factorization operators and second order symmetries}\label{symmetries}

As we have seen above, the action of the operators $L^{\pm}_{\l+m}$ on eigenfunctions of the Hamiltonian hierarchy change the label $\l+m$ in one unit, while keeping the energy. We can write this property as follows
\be
L^{+}_{\l+m}\Phi_{\l+m} (r)
\propto \Phi_{\l+m+1}  (r)\quad
L^{-}_{\l+m}\Phi_{\l+m+1} (r) 
\propto \Phi_{\l+m}  (r) .
\ee
If we recall that the label $\l+m$ is for the ${\cal J}_3$-eigenvalue, we can say that the operators $L^{\pm}_{\l+m}$ act essentially as lowering or raising operators for ${\cal J}_3$. However, when they act on the eigenstates of the Hamiltonians ${\cal H}_{\l}$ they preserve the energy eigenvalue.

Now, if we go back to the second order symmetries  ${\cal T}_1, {\cal T}_2$ of the initial Hamiltonian ${\cal H}$, and we introduce the operators ${\cal T}^{\pm}={\pm i}({\cal T}_1\mp  i\,{\cal T}_2)/2$, they satisfy the commutation rules
\be
[{\cal J}_3,{\cal T}^{\pm}] = \pm {\cal T}^{\pm} \qquad
[{\cal H}, {\cal T}^{\pm}] =0 \qquad [{\cal T}^+,{\cal T}^-] = -2\,  {\cal H} \, {\cal J}_3 .
\label{pm}
\ee
This means that ${\cal T}^{\pm}$ acting on the common eigenfunctions of ${\cal H}$ and ${\cal J}_3$ realises the same role played by $L^{\pm}_{\l+m}$: change the ${\cal J}_3$-eigenvalue in one unit while they leave that of ${\cal H}$ unaltered. Therefore, there must be a close relationship between both types of operators, as it was the case for the Coulomb problem \cite{lyman}. Indeed, if we write the eigenfunctions $\Psi_{\l}(r,\t)$ given in (\ref{23})--(\ref{23a}) in terms of $\Phi_{\l}(r)$ introduced in (\ref{ff}),
\be\label{psl}
\Psi_{\l}(r,\t) = r^{-1/2}\ Y_{\l}(\theta)\ \Phi_{\l}(r) 
\ee
and we use the first commutation relation in Eq.~(\ref{pm}), we have
\be
{\cal T}^{\pm}\Psi_{\l}(r,\t) 
\propto \Psi_{\l\pm1}(r,\t) .
\ee
In order to prove this relationship explicitly, let us use first (\ref{se1})--(\ref{se2}) and 
(\ref{hbar=1})--(\ref{secondordsym}) to express ${\cal T}^{\pm}$ in polar coordinates:
\be\label{spm}
\begin{array}{l}
\displaystyle
{\cal T}^+ = -\frac{i}2\, e^{i\,\t}\left\{ -2\left(\frac{i}{r}\p_{\t} +\p_r\right)
\left(\p_{\t} + \frac{i}{2}(\s_3+\s_0) \right) +
\left(\begin{array}{cc}
0&-e^{-i\,\t}\\
e^{i\,\t}&0
\end{array}\right)\right\}
\\[3.ex]
\displaystyle
{\cal T}^- = \frac{i}2\,\left\{ -2\left(-\frac{i}{r}\p_{\t} +\p_r\right)
\left(\p_{\t} + \frac{i}{2}(\s_3+\s_0) \right) -
\left(\begin{array}{cc}
0&-e^{-i\,\t}\\
e^{i\,\t}&0
\end{array}\right) \right\}e^{-i\,\t} .
\end{array}
\ee
Then, if we take into account that
\bea
&&\p_{\t}\Psi_{\l}(r,\t)= i\,  r^{-1/2}\ 
Y_{\l}(\theta) \left(\l \s_0 -\frac12\s_3\right)\, \Phi_{\l}(r) 
\label{1}\\[1.ex]
&&\p_{r}\Psi_{\l}(r,\t)= r^{-1/2}\ Y_{\l}(\theta)
\left(\frac{-1}{2r}+\p_r\right) \Phi_{\l}(r) 
\label{2}\\[1.ex]
&&
\left(\begin{array}{cc}
0&-e^{-i\,\t}\\
e^{i\,\t}&0
\end{array}\right)
\Psi_{\l}(r,\t)= (-i)\, r^{-1/2} \ Y_{\l}(\theta)
\s_2\, \Phi_{\l}(r) 
\label{3}
\eea
and we insert these identities in (\ref{spm}), we get the induced action of ${\cal T}^{\pm}$ on $\Phi_{\l}(r)$:
\be
{\cal T}^{\pm} \Phi_{\l}(r) = 
\left(\l+\frac12\right) \left[\mp\frac{d}{d r} +\frac{(\l\pm1/2) - (1/2)\,\s_3}{r}
-\frac{1/2}{\l\pm1/2} \, \s_2\right]  \Phi_{\l}(r) .
\ee
If we compare the above expression with the definition (\ref{not3}) of $L^{\pm}_{\l}$, we see that indeed, the action of the second order symmetries ${\cal T}^{\pm}$ induced in the space of the eigenfunctions $\Phi_{\l+m}(r)$, essentially coincide with the action of the intertwining operators $L^{\pm}_{\l+m}$, as we expected from the initial intuitive arguments. In fact, we get the normalized operators $\w L^{\pm}_{\l}$ defined in (\ref{norm}), and the commutation (\ref{pm}) of ${\cal T}^{\pm}$ induces in the space of functions $\Phi_{\l+m}(r)$ the relation displayed in (\ref{normalc}) between the operators $\w L^{\pm}_{\l+m}$.


\section{Ladder operators}\label{ladder}

In this section we will investigate the construction of ladder operators that will allow to link the excited states of a given Hamiltonian ${\cal H}_{\l}$.

Let us consider the eigenvalue equation (\ref{pauli}) for a given Hamiltonian ${\cal H}_{\l}$ corresponding to the energy $E_{\l}^n$ and eigenfunction $\Phi^n_{\l} (r)$ given in 
(\ref{exc})--(\ref{excenergy}). After multiplying by $r^2$, we reorder the terms in the following way:
\be
{\cal R}_{\l,n}\,\Phi^n_{\l} (r)\equiv\left[-r^2 \frac{d^2}{d r^2} -  r \s_2 + \frac{r^2}{4(\l+n+1/2)} - \l\, \s_3\right] \Phi^n_{\l} (r) = - \l^2 \Phi^n_{\l} (r) .
\ee
In  a similar way to what we did in (\ref{not1})--(\ref{not}), we can factorize the operators 
${\cal R}_{\l,n}$ in two ways
\be\label{qs}
{\cal R}_{\l,n} = Q^+_{\l,n} Q^-_{\l,n} + \om_{\l,n} =
Q^-_{\l,n+1} Q^+_{\l,n+1} + \om_{\l,n+1}  
\ee
where
\bea
&&Q^+_{\l,n}= \left[ -r\frac{d}{d r} + \frac{\l\, \s_3}{2(\l+n)} - (\l+n) +\frac12 + \frac{ \s_2}{2(\l+n+1/2)}\, r  \right] {\cal D}_{\l+n}^{-1}
\\[1.ex]
&&Q^-_{\l,n}= {\cal D}_{\l+n} \left[ r\frac{d}{d r} + \frac{\l\, \s_3}{2(\l+n)} - 
(\l+n) - \frac12 + \frac{  \s_2}{2(\l+n+1/2)}\, r  \right] 
\\[1.ex]
&&\om_{\l,n}= -\frac{\l^2}{4(\l+n)^2}- (\l+n)^2 +\frac14
\eea
and the operators ${\cal D}_{\l+n}$ are dilation operators defined by
\be
{\cal D}_{\l+n}\, r= \frac{\l+n+1/2}{\l+n-1/2 }\, r 
= \sqrt{\frac{E^n_\l }{E^{n-1}_\l }}\, r .
\ee
The operators $Q^{\pm}_{\l,n+1}$ act as intertwining of the diferential operators 
${\cal R}_{\l,n}$ and ${\cal R}_{\l,n+1}$:
\be
Q^+_{\l,n+1}\, {\cal R}_{\l,n} = {\cal R}_{\l,n+1}\, Q^+_{\l,n+1}\quad
Q^-_{\l,n+1}\, {\cal R}_{\l,n+1}= {\cal R}_{\l,n}\, Q^-_{\l,n+1}.
\ee
As a consequence, the set of operators $\{Q^{\pm}_{\l,n},\, n\in \Z^+\}$ will
act as lowering and raising operators for the Hamiltonian ${\cal H}_{\l}$,
\be
Q^+_{\l,n+1}: \Phi_{\l}^n \to \Phi_{\l}^{n+1}\qquad
Q^-_{\l,n+1}: \Phi_{\l}^{n+1} \to \Phi_{\l}^{n} .
\ee
From (\ref{qs}) we can compute the normalization of the eigenfunctions obtained in this way:
\be
||\Phi_{\l}^{n+1}||^2 = (\om_{\l,0}-\om_{\l,n+1})||\Phi_{\l}^n||^2 .
\ee 
In particular for $n=0$, we have $Q^-_{0}\, \Phi_{\l}^{0}= 0$. The operators $Q^{\pm}_{\l,n}$ constitute a non-trivial generalization of those well known for the Coulomb problem \cite{olmo}.

Finally, we can join the two families of `ladder operators' found up to now, $\{L^{\pm}_{\l+m}\}$ and $\{Q^{\pm}_{n}\}$, to build a `dynamical algebra' inside the hierarchy 
$\{ {\cal H}_{\l+m}\}$. However, only for the case $\l=1/2$ all these operators will connect exclusively physical states of the discrete spectrum.


\section{Remarks and conclusions}\label{conclusions}

In this paper we have studied a two-dimensional Pauli Hamiltonian,
which has two independent integrals of motion, from the point
of view of supersymmetric quantum mechanics. We have examined a series of properties that were not fully explored up to now:
\begin{itemize}
\item[(i)] 
We have considered some useful discrete symmetries since the very beginning. 
\item[(ii)] 
We also included in our study a family of Hamiltonian hierarchies labelled by the parameter $\l\in[0,1/2]$. For all these hierarchies the spectral problem was solved by means of the matrix shape-invariant method.

\item[(iii)] 
We have shown the relation between these intertwining operators and the second order symmetries. 
\item[(iv)] 
We also computed the ladder operators suitable to this matrix problem, which link the excited states of the same Hamiltonian.
\end{itemize}
In summary, we have shown that the methods employed along of this paper (SUSY quantum mechanics) constitute a very useful tool for the investigation of Pauli matrix Hamiltonians, and deserve to be exploited for more general situations.

\appendix

\section*{Appendix}
\setcounter{section}{1}

In this appendix we will show that the excited physical states $\Phi_\l^n(r)$ are well behaved near the origin and at infinity. We start with the expression (\ref{exc}) for the excited states, taking into account that the physical ground states are $\Phi_{\l+n}^0(r)\propto {\mathtt K}_{\l+n}(r)$, where ${\mathtt K}_{\l+n}(r)$ is given in  (\ref{sol1}).
For the sake of simplicity, we will use in the sequel the variable $a=\l+n$, and therefore (\ref{exc}) takes the form
\be\label{excA}
\Phi_{a-n}^n(r)=L_{a-n}^-L_{a-n+1}^- \cdots L_{a-2}^- L_{a-1}^-\, \Phi_{a}^0(r)
\ee
where, according to (\ref{sol1}), the ground state is given by
\be
\Phi_{a}^0(r) =   (2a+1)^{a+1}
\left(\begin{array}{c} z^{a+1}K_1  \left(z \right) \\
i\, z^{a+1} K_0 \left(z\right) \end{array}\right) 
\qquad z=\frac{r}{2a+1}.
\ee
Using the new variable $z$, the operators $L^-_b$ adopt the form
\be
\label{Lbb}
L^-_{b}= \frac1{2a+1} \left[
\frac{d}{d z} +\frac{2b+1 - \s_3}{2 z} -
\frac{2a+1}{2b+1}\ \s_2 \right] .
\ee
It is interesting to know the approximate form that have these operators near the origin ($r,z\approx 0$) and at infinity ($r,z\approx \infty$). Indeed, we have the following: 
\bea
  {\rm if } \quad r,z\approx 0, && \quad L^-_{b} \approx N^-_{b} = \frac1{2a+1} 
\left(\begin{array}{cc} \partial_z +\frac{b}z &0  \\
0 & \partial_z +\frac{b+1}z \end{array}\right) 
\\ [1ex]
{\rm if }\quad  r,z\approx \infty, && \quad L^-_{b} \approx G^-_{b} = \frac1{2a+1} 
\left(\begin{array}{cc} \partial_z  & i \frac{2a+1}{2b+1}   \\
-i \frac{2a+1}{2b+1} & \partial_z  \end{array}\right) .
\eea

\subsection{Behaviour near the origin}

In this approximation, we have that
\be
\Phi_{a-n}^n \approx N_{a-n}^- N_{a-n+1}^- \cdots N_{a-2}^-  N_{a-1}^-\, \Phi_{a}^0
\qquad n=1,2,\dots
\ee
and using
\bea
&& K_0(z) \approx \log (z) \qquad \quad K_1(z) \approx 1/{z} \label{k0k1cerca0}\\
&& K'_0(z)= -K_1(z) \qquad z K'_1(z)=-z K_0(z)-K_1(z)
\eea
we can prove that
\bea
 && N_{a-1}^-\, \Phi_{a}^0 \approx  (2a+1)^{a}
\left(\begin{array}{cc} 2a-1 & 0 \\
0 & 2a+1 \end{array}\right) 
\left(\begin{array}{c} z^{a}K_1 (z) \\
i\, z^{a} K_0(z) \end{array}\right) 
\nonumber
\\
 && N_{a-2}^-N_{a-1}^- \Phi_{a}^0 \approx  (2a+1)^{a-1}
\left(\begin{array}{cc} [2a-1][2a-3] & 0 \\
0 & [2a+1][2a-1] \end{array}\right) 
\left(\begin{array}{c} z^{a-1}K_1 (z) \\
i\, z^{a-1} K_0 (z) \end{array}\right) .
\nonumber
\eea
Therefore
\be 
\Phi_{a-n}^n \approx 2^n (2a+1)^{a-n+1}
\left(\begin{array}{cc} \frac{\Gamma(a+1/2)}{\Gamma(a+1/2-n)} & 0 \\
0 & \frac{\Gamma(a+3/2)}{\Gamma(a+3/2-n)} \end{array}\right) 
\left(\begin{array}{c} z^{a-n+1}K_1 (z) \\
i\, z^{a-n+1} K_0 (z) \end{array}\right) 
\nonumber
\ee
or
\be 
\nonumber
\Phi_{\l}^n \approx 2^n (2\l+2n+1)^{\l+1}
\left(\begin{array}{cc} \frac{\Gamma(\l+n+1/2)}{\Gamma(\l+1/2)} & 0 \\
0 & \frac{\Gamma(\l+n+3/2)}{\Gamma(\l+3/2)} \end{array}\right) 
\left(\begin{array}{c} z^{\l+1}K_1 (z) \\
i\, z^{\l+1} K_0 (z) \end{array}\right) .
\ee
From here, taking into account  (\ref{k0k1cerca0}), it is easy to conclude that for any 
$\l> 0$ the states $\Phi_{\l}^n$ are well behaved near the origin.

\subsection{Behaviour at the infinity}

For big values of the variables $z$ or $r$, we have that
\be
\Phi_{a-n}^n \approx G_{a-n}^- G_{a-n+1}^- \cdots G_{a-2}^-  G_{a-1}^-\, \Phi_{a}^0
\ee
and taking into account that in this region we can approximate
\be
K_\nu (z) \approx \sqrt{\frac{\pi}{2z}}\ e^{-z} \label{k0k1cercainfty}
\ee
after a simple calculation we get
\be
\Phi_{a-n}^n \approx \frac{ (-\s_2)^n}{(2a+1)^n}\ 
\frac{\Gamma(a-n+1/2)\Gamma(2a+1)}{\Gamma(a+1/2)\Gamma(2a-n+1)}\  \Phi_{a}^0
\ee
or
\be
\Phi_{\l}^n \approx \frac{ (-\s_2)^n}{(2\l+2n+1)^n}\ 
\frac{\Gamma(\l+1/2)\Gamma(2\l+2n+1)}{\Gamma(\l+n+1/2)\Gamma(2\l+n+1)}\ \Phi_{\l+n}^0.
\ee
Using  (\ref{k0k1cercainfty}), we conclude that for any $\l> 0$ the states $\Phi_{\l}^n$ are well behaved at infinity.

\section*{Acknowledgments}
This work has been partially supported by Spanish
Ministerio de Educaci\'on y Ciencia under Projects BMF2002-02000, BFM2002-03773, SAB2004-0143 (sabbatical grant of MVI), Ministerio de Asuntos Exteriores (AECI grant 0000147625 of SK) and Junta de Castilla y Le\'on (Excellence Project VA013C05). The research of MVI is also supported by the Russian grants RFFI 06-01-00186-a and RNP 2.1.1.1112. The authors would like to thank Prof. GP~Pron'ko for bringing this problem to their attention.



\end{document}